\documentclass[twoside]{dis08}
\usepackage[latin1]{inputenc}
\usepackage[dvips]{graphicx,epsfig,color}
\usepackage{wrapfig,rotating}
\usepackage{amssymb,amsmath,array}

\newcommand{\bfS}{\mbox{\boldmath $S$}}
\newcommand{\pup}{p^\uparrow}

\newcommand{\be}{\begin{equation}}
\newcommand{\ee}{\end{equation}}
\newcommand{\bea}{\begin{eqnarray}}
\newcommand{\eea}{\end{eqnarray}}
\newcommand{\bfk}{\mbox{\boldmath $k$}}

\def\kt{k_\perp}

\newcommand{\bfp}{\mbox{\boldmath $p$}}

\begin{document}
\title{The Sivers Function From SIDIS Data}
\author{M.~Anselmino$^1$, M.~Boglione$^1$, U.~D'Alesio$^{2,3}$, 
A.~Kotzinian$^4$,\\ S.~Melis$^1$\footnote{Talk presented by S. Melis at the 
XVI International Workshop on Deep-Inelastic Scattering and Related Subjects, 
DIS 2008, London}, F.~Murgia$^3$, {A.~Prokudin}$^{1,5}$, C.~T\"{u}rk$^1$
%
%
%
\vspace{.3cm}\\
%
1- Dipartimento di Fisica Teorica, Universit\`a di Torino and \\ 
          INFN, Sezione di Torino, Via P. Giuria 1, I-10125 Torino, Italy
\vspace{.1cm}\\
2- Dipartimento di Fisica, Universit\`a di Cagliari,\\ 
Cittadella Universitaria di Monserrato, I-09042 Monserrato (CA), Italy
\vspace{.1cm}\\
3- INFN, Sezione di Cagliari, C.P. 170, I-09042 Monserrato (CA), Italy
\vspace{.1cm}\\
4- CEA-Saclay, IRFU/Service de Physique Nucl\'eaire, 91191 Gif-sur-Yvette, 
France; \\ Yerevan Physics Institute, 375036 Yerevan, Armenia;
JINR, 141980 Dubna, Russia
\vspace{.1cm}\\
5- Di.S.T.A., Universit\`a del Piemonte Orientale
             ``A. Avogadro'', Alessandria, Italy
}

\maketitle

\begin{abstract}
We study the Sivers effect in transverse single spin asymmetries (SSA) for
pion and kaon production in Semi-Inclusive Deep Inelastic Scattering (SIDIS)  
processes. We perform a fit of $A^{\sin(\phi_h-\phi_S)}_{UT}$ 
taking into account the recent data
from HERMES and COMPASS Collaborations, which allow a new determination 
of the Sivers distribution functions for quark and anti-quark with $u$, $d$ 
and also $s$ flavours. Estimates for forthcoming SIDIS experiments at COMPASS 
and JLab are given.
\end{abstract}

Data on the transverse single spin asymmetry $A_{UT}^{\sin(\phi_{h}-\phi_S)}$ 
for polarized SIDIS processes,
$\ell \, p\,(\bfS) \to \ell ^\prime \, h \, X$, collected
by the HERMES~\cite{Airapetian:2004tw} and COMPASS~\cite{Alexakhin:2005iw}
Collaborations
allowed us~\cite{Anselmino:2005nn,Anselmino:2005ea} to perform a rather well 
constrained extraction of the 
Sivers distribution function \cite{Sivers:1989cc,Sivers:1990fh} for $u$ and 
$d$ quarks, assuming  
a negligibly small 
Sivers sea.
Recently, much higher statistics data 
on the $A_{UT}^{\sin(\phi_{h}-\phi_S)}$ azimuthal asymmetries for SIDIS have 
become available: in Ref.~\cite{Diefenthaler:2007rj} the HERMES Collaboration 
presents neutral pion and charged kaon azimuthal asymmetries, in addition 
to higher precision data on charged pion asymmetries; moreover,
Refs.~\cite{Martin:2007au,Alekseev:2008dn} show the COMPASS Collaboration 
measurements for separated charged pion and kaon asymmetries, 
together with some data for $K^0_S$ production. 

Here we present the analysis of these new experimental sets of 
data~\cite{Anselmino:2008sg}.
They give us a better understanding of the $u$ and $d$ flavour 
 Sivers distribution functions at low-intermediate $x$ and, most importantly, 
a first insight into the sea contributions to the Sivers functions.

The SIDIS transverse single spin asymmetry  $A^{\sin(\phi_h-\phi_S)}_{UT}$
is defined as  
\be
A^{\sin (\phi_h-\phi_S)}_{UT} = \label{def-siv-asym}
2 \, \frac{\int d\phi_S \, d\phi_h \,
[d\sigma^\uparrow - d\sigma^\downarrow] \, \sin(\phi_h-\phi_S)}
{\int d\phi_S \, d\phi_h \,
[d\sigma^\uparrow + d\sigma^\downarrow]}\,,
\ee
where $\phi_S$ and $\phi_h$ are the azimuthal angles identifying the
directions of the proton spin $\bfS$ and of the outgoing hadron $h$ in the 
$\gamma^{*}p$ c.m. frame, see Fig.~1 of Ref.~\cite{Anselmino:2008sg}.
Taking into account intrinsic parton motion, this 
transverse single spin asymmetry, can be written, at order $(\kt/Q)$, as:
\be
A^{\sin (\phi_h-\phi_S)}_{UT} \!= \!\label{hermesut}
\frac{\displaystyle  \sum_q \!\!\int
{\!\!d\phi_S \;\! d\phi_h \;\! d^2 \bfk _\perp}
\Delta^N \! f_{q/\pup} (x,\kt) \sin (\varphi -\phi_S)
\frac{d \hat\sigma ^{\ell q\to \ell q}}{dQ^2}
 D_q^h(z,p_\perp) \sin (\phi_h -\phi_S) }
{\displaystyle \sum_q \!\!\int {\!\!d\phi_S \,d\phi_h \, d^2 \bfk _\perp}\,
f_{q/p}(x,k _\perp) \frac{d \hat\sigma ^{\ell q\to \ell q}}{dQ^2}
 \; D_q^h(z,p_{\perp) }} \;\!\!,
\ee
where $\varphi$ defines the direction of the incoming
(and outgoing) quark transverse momentum,
$\bfk_\perp$ = $\kt(\cos\varphi, \sin\varphi,0)$;
$f_ {q/p}(x,\kt)$ is the unpolarized $x$ and $\kt$ dependent
parton distribution function (PDF);
$d \hat\sigma ^{\ell q\to \ell q}/dQ^2$ is the unpolarized
cross section for the elementary scattering \mbox{ $\ell q\to \ell q$}; 
$D_q^h(z,p_\perp)$ is the fragmentation function describing the 
hadronization of the final quark $q$ into the detected hadron $h$ with a 
light-cone momentum fraction $z$ and a transverse 
momentum $\bfp_\perp$ with respect 
to the fragmenting quark;
finally, $\Delta^N \! f_ {q/\pup}(x,k_\perp)$ is the Sivers function,
parameterized in terms of the unpolarized
distribution function as:
\be
\Delta^N \! f_ {q/\pup}(x,\kt) = 2 \, {\cal N}_q(x) \, h(\kt) \,
f_ {q/p} (x,\kt)\; , \label{sivfac}
\ee
with
\be
{\cal N}_q(x) =  N_q \, x^{\alpha_q}(1-x)^{\beta_q} \,
\frac{(\alpha_q+\beta_q)^{(\alpha_q+\beta_q)}}
{\alpha_q^{\alpha_q} \beta_q^{\beta_q}}\; ,
\quad\quad
h(\kt) = \sqrt{2e}\,\frac{k_\perp}{M_{1}}\,e^{-{k_\perp^2}/{M_{1}^2}}\; ,
\label{siverssxkt}
\ee
where $N_q\in[-1,1]$, $\alpha_q$, $\beta_q$ and $M_1$ (GeV/$c$) are free parameters
to be determined by fitting the experimental data. Notice that $h(\kt) \le 1$ for 
any $\kt$ and $|{\cal N}_q(x)| \le 1$ for any $x$, therefore the 
positivity bound for the Sivers function
is automatically fulfilled. For the unpolarized distribution and fragmentation 
functions, we adopt the common factorized gaussian form
\be
f_{q/p}(x,k_\perp) = f_q(x) \, \frac{1}{\pi \langle\kt^2\rangle} \,
e^{-{\kt^2}/{\langle\kt^2\rangle}}\,,
%
\quad\quad
%
D_q^h(z,p _\perp) = D_q^h(z) \, \frac{1}{\pi \langle p_\perp^2\rangle}
\, e^{-p_\perp^2/\langle p_\perp^2\rangle} \>,
\label{partonf}
\ee
with $\langle k_\perp^2\rangle=0.25  \;({\rm GeV}/c)^2 $ and $\langle p_\perp^2\rangle= 0.20 \;({\rm GeV}/c)^2$ 
fixed by analysing the 
Cahn effect in unpolarized SIDIS, as in Ref.~\cite{Anselmino:2005nn}.
%
%
The parton distribution functions $f_{q}(x)$ and the fragmentation 
functions $D_q^h(z)$ also depend on $Q^2$ via the usual QCD evolution,
which will be taken into account, at LO, in all our computations.

%
Fragmentation functions are a crucial ingredient of our fit.
We have considered three different sets: KRE~\cite{Kretzer:2000yf},
HKNS~\cite{Hirai:2007cx} and DSS~\cite{deFlorian:2007aj}.  
All these sets are basically equivalent as far as pion asymmetries are 
concerned.
However there are important differences in the description of kaon data.
In particular the DSS set, contrary to the other two sets,
is such that $D_{\bar s}^{K^+}(z)\gg D_{u}^{K^+}(z)$ over the whole $z$ range.
This feature is crucial when studying kaon production processes: first of all, 
it allows to reproduce kaon multiplicities at HERMES; secondly, it enables us 
to achieve kaon asymmetries larger than those corresponding to pion production.
For these reasons we have chosen the DSS set for our fit.
Contrary to the fragmentation sector, the use of different sets of unpolarized 
distribution functions does not affect our results significantly; here we use 
the GRV98LO set ~\cite{Gluck:1998xa}.
As the SIDIS data from HERMES and COMPASS have a limited coverage in $x$,
typically $x < 0.3-0.4$, the experimental asymmetries we are fitting contain
very little information on the large $x$ tail of the Sivers functions.
Therefore we assume the same value of $\beta$
(which is related to the shape of the Sivers functions at large $x$)
for all Sivers functions, setting 
$\beta_{sea} = \beta_{u} = \beta_{d} \equiv \beta$.
Notice that this choice artificially reduces the width of the uncertainty band at large $x$.
Moreover we assume the same $\alpha=\alpha_{sea}$ for all sea quarks.
Thus for this so called `broken sea' ansatz fit we then have 11 parameters.
\begin{table}[t]
\begin{center}
\begin{tabular}{|l|l|l|}
\hline
~&~&~\\
~~~$N_{u} = 0.35 ^{+ 0.08}_{-0.08} $ &
~~~$N_{d} = -0.90 ^{+ 0.43}_{-0.10} $ &
~~~$N_{s} = -0.24 ^{+ 0.62}_{-0.50} $~~ \\
~~~$N_{\bar u} =  0.04 ^{+ 0.22}_{-0.24} $&
~~~$N_{\bar d} =  -0.40 ^{+ 0.33}_{-0.44}$&
~~~$N_{\bar s} =  1 ^{+0}_{-0.0001}$ \\
~~~$\alpha _u = 0.73 ^{+ 0.72}_{-0.58}$ &
~~~$\alpha_d = 1.08 ^{+ 0.82}_{-0.65}$  &
~~~$\alpha_{sea} = 0.79 ^{+ 0.56}_{-0.47}$ \\
~~~$\beta = 3.46 ^{+ 4.87}_{-2.90}$ &
~~~$M_1^2 = 0.34 ^{+ 0.30}_{-0.16}$ (GeV/$c)^2$~  &
~~~~~ \\
~&~&~\\
\hline
\end{tabular}
\end{center}
\vspace{-0.3cm}
\caption{
Best values of the free parameters for the `broken sea' ansatz. 
The errors are determined according to the procedure explained 
in Appendix A of Ref.~\cite{Anselmino:2008sg}.
\label{fitpar_sivers}}
\end{table}
The results we obtain for these parameters by fitting simultaneously the four
experimental data sets on $A_{UT}^{\sin(\phi_h-\phi_S)}$, corresponding to
pion and kaon production at HERMES~\cite{Diefenthaler:2007rj} and
COMPASS~\cite{Martin:2007au}, are presented in Table~I
%
\begin{wrapfigure}{r}{0.65\columnwidth}
\centerline{\includegraphics[angle=-90,width=0.63\columnwidth]{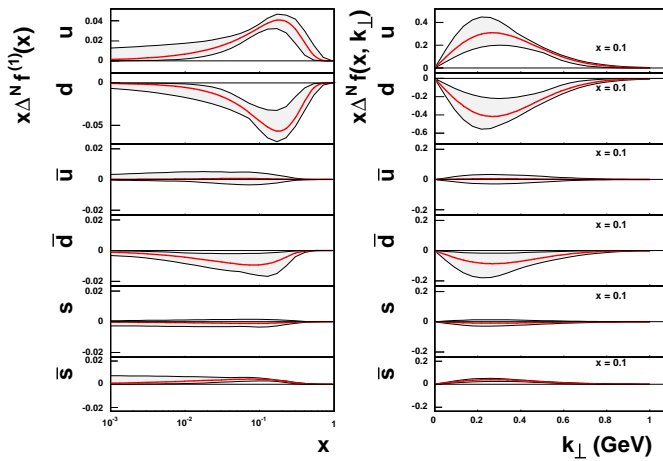}}
\caption{The Sivers distribution functions for $u$, $d$ and $s$ flavours as determined
by our simultaneous fit of HERMES and COMPASS data.
On the left panel, the first moment $x\,\Delta^N \! f^{(1)}(x)\equiv-x \,f_{1T}^{\perp (1) q}(x)$( 
see Eq.~17 of Ref.~\cite{Anselmino:2008sg}) is shown as a function of $x$ for each flavour, as indicated. 
Similarly, on the right panel, the Sivers distribution $x\,\Delta^N \! f(x,\kt)$ 
is shown as a function of $\kt$ at a fixed value of $x$ for each flavour, as 
indicated.}\label{Fig:MV}
\end{wrapfigure}
%
together with the corresponding errors, estimated according to the procedure
outlined in Appendix A of Ref.~\cite{Anselmino:2008sg}.

The fit performed under the `broken sea' 
ansatz shows a good description of pion and kaon asymmetries.
We obtained $\chi^2 = 1.20$ per data point for $K^+$ 
production at HERMES~\cite{Diefenthaler:2007rj}, while for pions we had 
$\chi^2 = 0.94$ per data point, for a total $\chi^2_{dof} = 1.00$.
Our results confirm that $\Delta^N \! f_{u/\pup} > 0$ and $\Delta^N \! f_{d/\pup} < 0$
as found in Ref.~\cite{Anselmino:2005ea}. Moreover HERMES data on kaon asymmetries cannot be 
explained without a sea-quark Sivers distribution. In particular we find that $\Delta^N \! f_{\bar s/\pup} > 0$.
Using the Sivers functions determined through our fit, we have given
predictions for $A_{UT}^{\sin(\phi_h-\phi_S)}$ for COMPASS 
experiment operating with a hydrogen target and 
at JLab, on proton, neutron and deuteron transversely polarized 
targets; for details see Ref.~\cite{Anselmino:2008sg}.

We have performed a comprehensive analysis of SIDIS data on Sivers azimuthal 
dependences.
It turns out that the data,
and in particular the unexpectedly large value of $A_{UT}^{\sin(\phi_h-\phi_S)}$ for $K^+$,
demand a non vanishing, and large, Sivers 
distribution for $\bar s$ quarks.
The other sea quark ($\bar u, \bar d, s$) 
contributions are less well determined, although they also 
seem to be non vanishing.

\begin{footnotesize}

\end{footnotesize}


\begin{thebibliography}{10}

\bibitem{Airapetian:2004tw}
HERMES, A.~Airapetian {\em et~al.},
\newblock Phys. Rev. Lett. {\bf 94}, 012002 (2005).

\bibitem{Alexakhin:2005iw}
COMPASS, V.~Y. Alexakhin {\em et~al.},
\newblock Phys. Rev. Lett. {\bf 94}, 202002 (2005).

\bibitem{Anselmino:2005nn}
M.~Anselmino {\em et~al.},
\newblock Phys. Rev. {\bf D71}, 074006 (2005).

\bibitem{Anselmino:2005ea}
M.~Anselmino {\em et~al.},
\newblock Phys. Rev. {\bf D72}, 094007 (2005).

\bibitem{Sivers:1989cc}
D.~W. Sivers,
\newblock Phys. Rev. {\bf D41}, 83 (1990).

\bibitem{Sivers:1990fh}
D.~W. Sivers,
\newblock Phys. Rev. {\bf D43}, 261 (1991).

\bibitem{Diefenthaler:2007rj}
HERMES, M.~Diefenthaler,
\newblock (2007),
\newblock arXiv:0706.2242 [hep-ex].

\bibitem{Martin:2007au}
COMPASS, A.~Martin,
\newblock Czech. J. Phys. {\bf 56}, F33 (2006).

\bibitem{Alekseev:2008dn}
COMPASS, M.~Alekseev {\em et~al.},
\newblock (2008),
\newblock arXiv:0802.2160 [hep-ex].

\bibitem{Anselmino:2008sg}
M.~Anselmino {\em et~al.},
\newblock (2008),
\newblock arXiv:0805.2677 [hep-ph].

\bibitem{Kretzer:2000yf}
S.~Kretzer,
\newblock Phys. Rev. {\bf D62}, 054001 (2000).

\bibitem{Hirai:2007cx}
M.~Hirai, S.~Kumano, T.~H. Nagai, and K.~Sudoh,
\newblock Phys. Rev. {\bf D75}, 094009 (2007).

\bibitem{deFlorian:2007aj}
D.~de~Florian, R.~Sassot, and M.~Stratmann,
\newblock Phys. Rev. {\bf D75}, 114010 (2007).

\bibitem{Gluck:1998xa}
M.~Gluck, E.~Reya, and A.~Vogt,
\newblock Eur. Phys. J. {\bf C5}, 461 (1998).

\end{thebibliography}
\end{document}